\documentclass[a4paper]{article}

\usepackage{INTERSPEECH2020}
\usepackage{graphicx,amsmath}
\usepackage{multirow}
\usepackage{rotating}
\usepackage{makecell}
\usepackage{booktabs}
\usepackage{adjustbox}
\usepackage[table,xcdraw]{xcolor}

\title{Sub-band Knowledge Distillation Framework for Speech Enhancement}
\name{Xiang Hao$^{1,2,\dagger}$, Shixue Wen$^{2,\dagger}$\footnotemark[2], Xiangdong Su$^{1*}$\footnotemark[1], Yun Liu$^2$, Guanglai Gao$^{1}$ and Xiaofei Li$^3$}
\address{
  $^1$College of Computer Science, Inner Mongolia University, Hohhot, China \\
  $^2$AI Interaction Division, Sogou Inc., Beijing, China\\
  $^3$Westlake University, Hangzhou, China
  }
\email{haoxiangsnr@gmail.com}

\begin{document}

\maketitle


\begin{abstract}
    In single-channel speech enhancement, methods based on full-band spectral features have been widely studied. However, only a few methods pay attention to non-full-band spectral features. In this paper, we explore a knowledge distillation framework based on sub-band spectral mapping for single-channel speech enhancement.
    Specifically, we divide the full frequency band into multiple sub-bands and pre-train an elite-level sub-band enhancement model (teacher model) for each sub-band. These teacher models are dedicated to processing their own sub-bands. Next, under the teacher models' guidance, we train a general sub-band enhancement model (student model) that works for all sub-bands. Without increasing the number of model parameters and computational complexity, the student model's performance is further improved. To evaluate our proposed method, we conducted a large number of experiments on an open-source data set. The final experimental results show that the guidance from the elite-level teacher models dramatically improves the student model's performance, which exceeds the full-band model by employing fewer parameters.
\end{abstract}

\renewcommand{\thefootnote}{\fnsymbol{footnote}}
\footnotetext[2]{Xiang Hao and Shixue Wen equally contributed to this work.}
\footnotetext[1]{Xiangdong Su is the corresponding author.}

\section{Introduction}

In recent years, single-channel speech enhancement methods based on deep learning have been proposed in large numbers, and they can be divided into time domain and frequency domain methods. Time-domain methods~\cite{segan}~\cite{Hao2019}~\cite{tcnn}~\cite{wavenet_d} typically use the time-domain signal of noisy speech as a direct input to a neural network, and the learning target is the clean speech signal. Frequency-domain methods~\cite{phase_sensitive_bilstm}~\cite{regression_approch}~\cite{se_overview}~\cite{ccrn} typically use the spectral features of noisy speech (magnitude spectrum, phase spectrum, complex spectrum, etc.) as inputs to a neural network, and learning targets are the spectral features of clean speech or some masks (e.g., ideal binary mask~\cite{ibm_as_goal_wang_2005}, ideal ratio mask~\cite{irm}, complex ideal ratio mask~\cite{cIRM}). The frequency-domain methods still dominate the vast majority of current single-channel speech enhancement methods due to the high dimension and the lack of apparent geometric structure for the time domain signal.

In the frequency domain, most speech enhancement methods focus on the full-band spectral representation. They take the full-band feature sequence directly as the input of the neural network.
Unlike them, this paper proposes a sub-band speech enhancement method. This method uses the magnitude feature sequence within a certain sub-band on the noisy spectrogram as the input of the model. The learning target is the magnitude feature sequence of the corresponding sub-band of the clean speech.
This method has some advantages over the full-band methods.
\textbf{(1)} For neural networks, the simpler the features to be learned, the easier the optimization of the network. Spectral features are highly patterned features, with low, medium, and high frequencies having significantly different data distributions. The reasonable division of sub-bands is equivalent to introducing a priori knowledge to the model so that the model can focus on learning more stable features on each sub-band. This method is similar to the partition technology in face recognition. The human face has a fixed pattern. By only hard segmenting the human face (eg., mouth, eye.), and then independently identifying each segment of the face and integrating them, the recognition accuracy can be improved~\cite{face_partition}.
\textbf{(2)} Most speech enhancement methods use an L-norm ($\ell_1, \ell_2$) loss function, for which the contribution of each frequency band to the total loss is identical. This may not be suitable for the actual situation that the speech energy is mainly concentrated in the low and medium frequencies. Correspondingly, the low and medium frequencies play a more important role in the loss. However, high-frequency information is essential for perceptual evaluation. The proposed method optimizes and calculates the loss for each sub-band separately to avoid the problem of applying the L-norm loss to the full-band spectra. Besides, a proper division of sub-bands also allows the model not to lose the ability to capture cross-band information too much.
Some of the previous works have been exploring non-full-band methods.
In~\cite{li_multichannel_2019}, a narrow-band LSTM, is proposed to deal with multichannel speech enhancement. The LSTM takes as input a sequence of TF bins associated with a frequency bin of multichannel noisy speech signals. The output of the model is a sequence of magnitude ratio masks at the same frequency bin. Experimental results show that the narrow-band method has outstanding performance.
In~\cite{wei_single-channel_2018}, the spectrogram is divided into different non-uniform sub-bands, and spectral subtraction is performed separately in each sub-band to obtain accurate noise estimates.
In~\cite{wang_exploring_2013}, the sub-band feature is used to classify speech or noise, which indicates that the local spectral pattern is informative for discriminating between speech and other signals.

Different sub-bands have very different data distributions, and it is undoubtedly challenging to learn them simultaneously through a single small-scale model. The integration using multiple sub-models is an effective method, but it is undoubtedly troublesome during deployment.
In this paper, we extend the sub-band enhancement model to a knowledge distillation framework containing multiple elite-level teacher models and one general-purpose student model for speech enhancement.
Knowledge distillation was first proposed by~\cite{hinton2015distilling}. As a compression~\cite{model_compression_distilling} or training technique~\cite{On_the_Efficacy_of_Knowledge_Distillation}~\cite{Towards_Understanding_Knowledge_Distillation}, it has been widely used in machine translation~\cite{nmt_knowledge_distillation}, image object detection~\cite{object_detection_knowledge_distillation}, speech recognition~\cite{danpovey_distilling}~\cite{speech_recogition_2016distilling} and other fields.
Knowledge distillation usually uses the category probability labels of a large teacher model and original labels as the learning goal of a small student model. It makes the small student model have the learning ability close to the large teacher model.
In the proposed sub-band knowledge distillation framework, we first trained multiple sub-band enhancement models (teacher models). These models specialize in feature mapping in a certain sub-band, and they can well conduct the speech enhancement in their sub-bands. After that, we use these teacher models to guide a sub-band enhanced model (student model) to learn all sub-bands. Finally, without increasing the number of parameters and the computational cost of the student model, the student model's performance is further improved.

In order to evaluate the effectiveness and performance of the proposed method, we conducted several experiments and analyses on an open-source data set.
The final experimental results show that the teacher model's guidance dramatically improves the student model's performance. Moreover, the performance of the student model exceeds the corresponding full-band model.

\section{Method}

We use the representation of the speech signal in the short-time fourier transform (STFT) domain:
\begin{equation}
    X(t,f) = S(t,f) + N(t,f)
\end{equation}
where $X(t,f)$, $S(t,f)$ and $N(t,f)$ respectively represent the time-frequency (T-F) bin of noisy speech, clean speech and interference noise at time frame $t$ and frequency bin $f$. $t \in Z^+, 1 \leq t \leq T$. $f \in Z^+, 0 \leq f \leq F - 1$. $T$ and $F$ denote the total number of frames and the total frequency bins expressed in the frequency domain, respectively. For the closed interval $[0, F - 1]$, we take an ordered sequence of equally spaced points:
\begin{eqnarray}
    0 = f_0 < f_1 < f_2 < ... < f_n = F - 1
\end{eqnarray}
We refer to the closed interval $[f_i,f_{(i+1)}]$ as the sub-band, where $0 \leq i \leq n - 1$. $[0, F - 1]$ contains $n$ sub-bands. The frequency width of all sub-bands is fixed at $\lfloor \frac{F}{n} \rfloor$ except for the last one. In this paper, we use the magnitude component of the speech in the STFT domain as the input and training target. We use a single neural model to map all sub-bands:
\begin{gather}
    |\hat{S}(i)| = G(|X(i)|) \\
    |X(i)| = \left(
    \begin{smallmatrix}
            |X(1, f_i)|    &  |X(2, f_i)|    & \cdots        &  |X(T, f_i)|   \\
            |X(1, f_{i} + 1)|    & |X(2, f_{i} + 1)|   & \cdots        &|X(T, f_{i} + 1)|   \\
            \vdots       & \vdots       & \ddots       & \vdots       \\
            |X(1, f_{i + 1})| & |X(2,f_{i + 1})| & \cdots & |X(T, f_{i + 1})|
        \end{smallmatrix}
    \right) \notag
\end{gather}
where $|X(i)|, 0 \leq i \leq n - 1$ is the noisy magnitude features in the $i$-th sub-band, $G( \cdot )$ is the sub-band enhancement model, and $|\hat{S}(i)|$ is the enhanced feature in the $i$-th sub-band with the same dimension as $|X(i)|$. We treat the divided sub-bands as independent units, and the model can be optimized for each sub-band individually. In an extreme case, we can divide $F$ sub-bands.

Considering that the model simultaneously handles sub-bands with different data distributions is very challenging, inspired by~\cite{hinton2015distilling}, we introduce the sub-band enhancement model into a knowledge distillation-based framework. The basic principle of knowledge distillation is that a soft label generated by the large model contains more information than the original target label. In general, if a large model gives a higher probability of the input features for specific categories, it means that those categories are similar to the real ones. We typically use knowledge distillation to distill the knowledge contained in a large-scale model (teacher model) into a smaller-scale model (student model) to compress model or to enhance performance.

In this paper, our knowledge distillation framework contains multiple teacher models for local sub-bands and a student model for all sub-bands. We train these teacher models so that they have excellent performance in their specific sub-band. During the student model training, we will select different teacher models according to the input sub-band to guide the student model, improving the student model's performance without changing the original student model scale.

\subsection{Training of the teacher models}

In order to build a sub-band knowledge distillation framework, we construct $n$ sub-band enhancement models $G_0 (\cdot), G_1(\cdot), ..., G_{(n-1)} (\cdot)$ to map $n$ magnitude sub-bands respectively:
\begin{gather}
    |\hat{S}| = \left(
    \begin{matrix}
            G_0(|X(0)|) \\
            G_1(|X(1)|) \\
            \vdots      \\
            G_{n-1}(|X(n-1)|)
        \end{matrix}
    \right)
\end{gather}
Here, the $|\hat{S}|$ is the magnitude spectra enhanced by $n$ different sub-band enhancement models. Each sub-band enhancement model is only responsible for enhancing frequency bands within a fixed range, and they are experts in the frequency bands that belong to them. We call this $N$ band enhancement models as teacher models.
When these teacher models are thoroughly trained, we fixed their weights. Subsequently, we used these teacher models to guide a sub-band enhancement model (student model) to learn different sub-bands better.

\subsection{Distillation}

When we select any one of the $n$ sub-bands to feed into the student model, we use the corresponding teacher model to guide the training of the student model. The training objective of the student model is the clean speech and the enhanced speech of the teacher model. We define the loss function $L$ as follows.
\begin{eqnarray}
    \label{eq:loss}
    L=(G_s (|X_i |)-|S_i |)^2 + \alpha (G_s (|X_i |)-G_i (|X_i |) )^2
\end{eqnarray}
where $G_s$ is the student sub-band enhancement model, and $G_i$ is the teacher model for the $i$-th sub-band, whose weights are fixed when training the student model. $|S_i|$ is the $i$-th sub-band corresponding to the clean speech, and $\alpha$ is the weight, representing the proportion of the loss related to the teacher model in the total loss.
It is worth mentioning that to ensure that the student model can get enough beneficial information from the teacher model, we usually train the teacher model with a model size greater than or equal to the student model.

\section{Experiment}

\subsection{Dataset and metrics}
In this paper, we evaluate our proposed method using an open-source per-mixed data set\footnote[1]{http://dx.doi.org/10.7488/ds/1356} that contains a large number of noises and speakers.
This data set includes 30 randomly selected speakers from Voice Bank Corpus~\cite{voice_bank_corpus}. Each speaker contains approximately 400 sentences.
Twenty-eight speakers (14 males and 14 females) were used for training, and two speakers (1 male and 1 female) for testing.
Twenty-eight speakers were used for training and two speakers for testing.

The training set of this data set contains a total of 10 different noises (2 synthetic noises and 8 noises selected from the Demand data set~\cite{demand}) and 4 different SNRs (0, 5,10, and 15dB). In total, the training set contains 40 different noise conditions (10 noises $\times$ 4 SNRs). Each training speaker has about 10 sentences in each noise condition.
The test set contains a total of 5 noise types (selected from the Demand data set~\cite{demand}) not present in the training set and 4 slightly higher SNRs (2.5, 7.5, 12.5, and 17.5dB) than the training set.
In total, the test set contains 20 different noise conditions (5 noises $\times$ 4 SNRs). There are around 20 different sentences in each condition per test speaker.
There is no validation set in the original data set to evaluate the training process, and we randomly selected 1000 speeches from the training set as a validation set.

In this paper, we use the perceptual evaluation of speech quality (PESQ)~\cite{pesq} and the short-term objective intelligibility (STOI)~\cite{stoi} to evaluate the quality and intelligibility of speech, respectively. They have been used in a large number of speech enhancement experiments.

\subsection{Implementation and details}

The student model and teacher models use the same model structure.
We stacked two bidirectional long short-term memory (LSTM) layers and one fully connected layer. We applied ReLU as an output activation layer.
The number of memory cells in each layer of LSTM will be set according to specific experiments.
The sampling rate of all speeches is 16000 Hz.
For the input of the model, 161-dimensional magnitude spectral features were extracted by performing STFT using a 320-point Hann window with an overlap of 50\%.

We trained the teacher models ($G_0$, $G_1$, ..., $G_{n-1}$) one by one.
The sub-band processed by the $G_0$ model is the lowest one (low frequency) on the spectrogram, and the sub-band processed by the $G_{n-1}$ model is the highest one (high frequency) on the spectrogram.
Each teacher model handles its own sub-band with a size of  $\lfloor \frac{161}{n} \rfloor$.
For the few high-frequency bands remaining in the spectrogram, we do not use them.
The student model needs to deal with all sub-bands. Before each training epoch starts, we randomly select any one of the n sub-bands as input to the student model.
In the inference stage, we use the trained student model to enhance the sub-bands of the noisy speech one by one. For the few high-frequency bands remaining in the noisy spectrogram, we do not enhance them. 
Then, combining with the phase component of the noisy STFT spectrogram, we use the inverse STFT to convert the enhanced magnitude spectrogram into the time domain speech waveform.
We shuffle the data set before each epoch begins by training a student model or a teacher model.

All hyperparameters are the same for all models.
We use mean square error (MSE, $\ell_2$) as the loss function.
We use Adam optimizer~\cite{adam} (decay rate $\beta_{1} = 0.9$, $\beta_{2} = 0.999$) and set the batch size to 600.
We set the initial learning rate of the models to a small constant of 0.0002.
When the loss of the model on the training set is oscillating, we appropriately reduce the learning rate.
When the loss of the model on the validation set does not drop noticeably for many epochs, we stopped training.
According to the preliminary experimental results on the current validation set, we set $\alpha$ in Equation~(\ref{eq:loss}) to $0.1$.

\subsection{Result}

\subsubsection{Effect of different sub-band sizes on model performance}

In our proposed method, each sub-band requires a teacher model. A question that must be answered is which is the proper size of the sub-band.
We trained three sets of sub-band enhancement models of different scales. The three sets of models have the same model structure except for the difference in the number of memory cells per layer in LSTM (256, 512, and 1024). Each set contains four sub-band enhancement models, and their inputs are all sub-bands. The size of the sub-bands is 1, 20, 40, and 80, respectively. In total, we trained 12 sub-band enhancement models.
We evaluate the performance of these 12 models on the test set, and the results are listed in Table~\ref{tab:differentSubBandLength}.


From Table~\ref{tab:differentSubBandLength}, we noticed that regardless of the size of the model and the sub-band size, the enhanced speech quality and intelligibility have been prominently improved compared to the original noisy speech.
When the size of the sub-band is 1, the input of the model is a single frequency band. We usually call this kind of model a narrow-band model. From the table, we can note that regardless of the scale of the model, the narrow-band model is the worst among all control groups. This is because the narrow-band model has very little frequency context information, and it is impossible to explore the feature information across the frequency band.

We also notice that as the size of the sub-band increases, the quality and intelligibility of the enhanced speech increase overall. This indicates that the large-sized sub-band is more conducive to the model to explore the local feature across the band. We also noticed that when the sub-band length is 40, the quality and intelligibility of the enhanced speech are slightly better than when the sub-band length is 80. This possibly indicates that for most of the local features, the sub-band size of 40 is already enough to cover them. We will use the sub-band enhancement models with an input size of 40 as our basic models.

\begin{table}[h]
    \centering
    \caption{Comparison of the models of different scales under different sizes of sub-bands.}
    \label{tab:differentSubBandLength}
    \setlength{\tabcolsep}{1.5mm}{
        \begin{tabular}{@{}ccccccc@{}}
            \toprule
                  & \multicolumn{3}{c}{PESQ}  & \multicolumn{3}{c}{STOI (\%)}                                                                        \\ \cmidrule{2-4} \cmidrule{5-7}
            Model & 256                       & 512                           & 1024           & 256             & 512             & 1024            \\ \midrule
            Noisy & \multicolumn{3}{c}{1.971} & \multicolumn{3}{c}{92.106}                                                                           \\
            1     & 2.151                     & 2.224                         & 2.277          & 92.415          & 92.174          & 93.082          \\
            20    & 2.369                     & 2.430                         & 2.441          & 92.877          & 93.265          & 93.476          \\
            40    & \textbf{2.425}            & \textbf{2.497}                & 2.512          & \textbf{93.421} & \textbf{93.540} & 93.721          \\
            80    & 2.417                     & 2.490                         & \textbf{2.513} & 93.243          & 93.471          & \textbf{93.727} \\ \bottomrule
        \end{tabular}
    }
\end{table}

\begin{table*}[t]
    \centering
    \caption{The mean square error (MSE, $\ell_2$ loss) of the teacher models and the student models (without the guidance of the teacher models) with the different number of memory cells per layer in LSTM for each sub-band on the test set. C is the number of memory cells per layer in LSTM.}
    \label{tab:mse}
    \begin{tabular}{@{}ccccccccc@{}}
        \toprule
             & \multicolumn{4}{c}{Student Model} & \multicolumn{4}{c}{Teacher Model}                                                                                                                            \\ \cmidrule(l){2-5} \cmidrule(l){6-9}
        C    & 0-40                              & 40-80 ($10^{-4}$)                 & 80-120 ($10^{-4}$) & 120-160 ($10^{-4}$) & 0-40           & 40-80 ($10^{-4}$) & 80-120 ($10^{-4}$) & 120-160 ($10^{-4}$) \\ \midrule
        256  & 0.036                             & 10.027                            & 3.647              & 2.801               & 0.031          & 9.230             & 3.578              & 2.204               \\
        512  & 0.026                             & 9.701                             & 3.594              & 2.508               & 0.019          & 9.038             & 3.494              & 1.792               \\
        1024 & 0.025                             & 9.162                             & 3.295              & 2.095               & \textbf{0.013} & \textbf{8.716}    & \textbf{3.020}     & \textbf{1.634}      \\ \bottomrule
    \end{tabular}
\end{table*}

\subsubsection{Comparison between teacher models and student models}
Table~\ref{tab:mse} comprehensively compares the $\ell_2$ loss between the teacher models and the student models (without the guidance of the teacher models) with the different number of memory cells per layer in LSTM for each sub-band. C is the number of memory cells. The left half of the table is the result area of the student models, and the right half is the result area of the teacher models.
For example, in the result area of the student models, the value in the first column (0 to 40) of the first row (C is 256) shows the result of the student model with 256 memory cells and for the 0 to 40 sub-band. The input to the student model is one of the four sub-bands in the training sample (random reselection for each iteration). 
The training target of the student model is solely the corresponding sub-band of the clean speech. The value of $0.036$ represents the $\ell_2$ loss between the enhanced test sample and the target.
In the result area of the teacher models, the first column (0 to 40) of the first row (C is 256) lists the teacher model result with 64 memory cells and for the 0 to 40  sub-band. 
The training data is a sub-band with a frequency range of 0 to 40 in each sample, and the target is the corresponding sub-band of the clean speech. The value of $0.031$ is the $\ell_2$ loss between the enhanced test sample and the target.
$10^{-4}$ in the table means that the value in the table is multiplied by $10^{-4}$.
In total, we trained 3 student models (3 different numbers of memory cells) and 12 teacher models (3 different numbers of memory cells $\times$ 4 sub-bands).


Observing the results in the table, we noticed some phenomena.
\textbf{(1)} It is not surprising that the teacher models trained for specific sub-bands are better than the student model trained for all sub-bands. This shows that although the student model can reduce the pressure of learning by capturing the commonality of each sub-band, the benefits of this commonality are not as good as the integration of multiple dedicated models.
\textbf{(2)} We also notice that as the number of memory cells increases, the $\ell_2$ loss of the teacher model and the student model on the test set is decreasing overall. This is as expected.
\textbf{(3)} In the 0 to 40 sub-band, the $\ell_2$ loss is much higher than that of 40 to 160. This is mainly because human speech energy is concentrated primarily in the low-frequency part of the spectrogram, resulting in a complicated data distribution. This also shows that the different frequency bands in the spectrogram contribute differently to the speech quality and intelligibility.

\subsubsection{Effectiveness of teacher model guidance}
From the above 12 teacher models, we select the 4 teacher models that perform best in the 4 sub-bands, which will be used to guide the training of student models.
Considering that we generally distill the knowledge of large models into small models in knowledge distillation, we set up two sets of models. These two sets contain 256 memory cells and 512 memory cells per layer in LSTM, respectively. Each set includes a full-band model (F), a student model without teacher model guidance (S1), and a student model with teacher model guidance (S2).
We list the performance comparison results of these models in Table 3.

In Table~\ref{tab:comparisonTeachers}, S1 model and S2 model have the same number of parameters. Since the input of the F model is the full-band features, the number of parameters of the F model is higher than that of the S1 model and S2 model.
Whether the number of memory cells is 256 or 512, we can find some similar conclusions.
\textbf{(1)} The performance of the S1 model is slightly worse than the F model. This may be because full-band input can bring more context information, which can help the model to capture the features on the spectrogram better. However, the performance gap is tiny, only 0.56 to 0.66\% and 0.29 to 0.30\% higher on PESQ and STOI, respectively. This may be because the S1 model can cover 40 frequency bands at a time, which is fairly sufficient to capture most of the local feature information. On the other hand, S1 has a 7\% reduction in the parameter amount compared to the F model. Considering that the speech enhancement model is usually deployed offline on hardware, this slight performance degradation is normally acceptable.
\textbf{(2)} With the guidance of the elite teacher model, the S2 model achieved better results than S1, which shows that the supervision of the teacher models is effective. It is worth mentioning that this improvement does not increase the number of parameters and the computational cost of the model. We also note that although the S2 model also does not have complete full-band context information, its performance noticeably exceeds model F.


\begin{table}[]
    \centering
    \caption{Demonstrates the effectiveness of the sub-band knowledge distillation framework. C represents the number of memory cells per layer in LSTM. \#Params represents the number of parameters of the model.}
    \label{tab:comparisonTeachers}
    \begin{tabular}{cccccc}
        \toprule
        Model & C   & \#Params (M) & PESQ           & STOI (\%)       \\ \midrule
        Noisy & -   & -            & 1.971          & 92.106          \\ \midrule
        F     & 256 & 2.52         & 2.420          & 93.412          \\
        S1    & 256 & 2.21         & 2.404          & 93.133          \\
        S2    & 256 & 2.21         & \textbf{2.471} & \textbf{93.751} \\ \midrule
        F     & 512 & 9.23         & 2.511          & 93.817          \\
        S1    & 512 & 8.61         & 2.497          & 93.540          \\
        S2    & 512 & 8.61         & \textbf{2.563} & \textbf{94.129} \\ \bottomrule
    \end{tabular}
\end{table}

\section{Conclusion}

In this paper, we propose a sub-band knowledge distillation framework for single-channel speech enhancement. We divide the spectrogram into multiple sub-bands, and train elite-level sub-band enhancement models (teacher models) on each sub-band. These models are then used to guide the training of general sub-band enhancement models (student model).
We evaluated our proposed method on an open-source data set.
We found that compared to the full-band model, although the sub-band enhancement model loses the ability to capture global cross-band information, the performance does not show a significant drop as expected. We also found that the teacher models further improves the performance of the sub-band enhancement model. Moreover, the performance of the student model exceeds the corresponding full-band model.

\bibliographystyle{IEEEtran}
\bibliography{mybib}
\end{document}